\documentstyle[prc,aps,twocolumn,psfig,epsfig]{revtex}

\input epsf

\def\Journal#1#2#3#4{{#1} {\bf #2}, #3 (#4)}

\def\NPB{{\em Nucl.~Phys.} B}
\def\NPA{{\em Nucl.~Phys.} A}
\def\PLB{{\em Phys.~Lett.}  B}
\def\PRL{\em Phys.~Rev.~Lett.~}
\def\PRD{{\em Phys.~Rev.} D}
\def\PRC{{\em Phys.~Rev.} C}
\def\ZPC{{\em Z.~Phys.} A}
\def\ZPC{{\em Z.~Phys.} C}

\def\EPJC{{\em Eur.~Phys.~J.} C}
\def\PRep{\em Phys.~Rep.~}

\newcommand{\expl}{\langle \!\langle}
\newcommand{\expr}{\rangle \!\rangle}

\begin{document}
\draft


\title{Antihyperon-Production in Relativistic Heavy Ion Collisions}

\author{
Carsten Greiner\footnote{E-mail: Carsten.Greiner@theo.physik.uni-giessen.de}
and
Stefan Leupold\footnote{E-mail: Stefan.Leupold@theo.physik.uni-giessen.de}
}

\address{
Institut f\"ur Theoretische Physik, Justus-Liebig--Universit\"at
Giessen,\\
Heinrich-Buff-Ring 16, D-35392 Giessen, Germany
}

\date{September 2000}

\maketitle

\begin{abstract}

Recently it has been shown that the observed antiproton yield in heavy-ion collisions
at CERN-SpS energies can be understood by multi-pionic interactions like
$\pi \pi \pi \pi \pi \leftrightarrow p \bar{p}$ which enforce local
chemical equilibrium of the antiprotons with the nucleons and pions.
Here we show that antihyperons are driven
towards local chemical equilibrium with pions, nucleons and kaons
on a timescale of less than 3 fm/c
when applying a similar argument for the antihyperons
by considering the inverse channel of annihilation reactions like
$\bar{Y}+p \leftrightarrow n_1\pi + n_2 K$.
These multi-mesonic reactions easily explain the
antihyperon yields at CERN-SpS energies
as advertised in pure thermal, hadronic models
without the need of a quark gluon plasma phase. In addition, the argument
also applies for AGS energies.

\end{abstract}

%
%

The prime focus for present and future ultrarelativistic heavy ion collisions
is to study the collective behavior of nuclear or hadronic matter at extreme
conditions like very high temperatures and energy densities.
A particular goal lies in the identification of a new state
of matter formed in central collisions, the quark gluon plasma (QGP),
where quarks and gluons are deliberated
from the nucleons and move freely over an extended, macroscopically large
space-time region. Indeed, a couple of crucial
observations obtained by the various experimental groups
within the Lead Beam Programme at the CERN-SpS
are thought to be not well described 
by hadronic models but indicate strong `circumstantial evidence' for
the formation of the QGP \cite{HJ00}.

As one of the major diagnostic probes for the
short-time existence of a QGP, measurements of strange and anti-strange
particles at CERN-SpS
by the WA97 Collaboration \cite{WA97a}
and the NA49 Collaboration \cite{NA49} have received much attention \cite{SQM98}.
(Anti-)strangeness enhancement relative to pp-data has been predicted
already a long time ago as a QGP signal: Especially the
antihyperons as well as the multi-strange baryons
were advocated as the appropriate candidates \cite{KMR86}.
It was argued that it would take
too much time compared
to the estimated overall collision time
for these special strange particles to come
close to their chemical equilibrium values within a thermalized
fireball environment of hadronic particles. For example
the strange antibaryons are thought
to be produced either by - highly threshold suppressed -
binary strangeness production reactions like e.g.
\begin{eqnarray}
\pi + \bar{p} & \rightarrow & \bar{K} + \bar{\Lambda }, \bar{\Sigma } 
\nonumber \\ & & \label{sproda}  \\
\pi + \bar{\Lambda } & \rightarrow & \bar{K} + \bar{\Xi } \nonumber 
\end{eqnarray}
or, successively, by further binary strangeness exchange reactions with the
(to be produced) kaons like e.g.
\begin{eqnarray}
K + \bar{p} & \rightarrow & \pi + \bar{\Lambda }, \bar{\Sigma } \nonumber
\\ & & \label{sprodb}  \\
K + \bar{\Lambda } & \rightarrow & \pi + \bar{\Xi }   \nonumber
\end{eqnarray}
with rather low cross sections.
Antihyperons should indeed be very rare and exotic probes.
On the other hand, assuming the
existence of a temporarily present phase of QGP, the rather
light strange quarks, following simple kinetic arguments,
can be produced much more abundantly
by gluon fusion. Together with the light quarks
and anti-quarks, the strange quarks
can now easily be redistributed into the to be produced hadrons.
By a simple coalescence estimate for the hadronization the QGP
can account for the antihyperons to come close to their
chemical equilibrium values
immediately at the onset of the hadronic phase \cite{KMR86}.

Experimentally, and most strikingly, on the other hand,
the relative enhancement to pp-collisions
is found to increase with the strange quark content of the produced
hadrons \cite{WA97a,NA49}. This fact is in strong conflict with
a hadronic rescattering scenario if the needed secondary production
of additional antihyperons (or multi-strange baryons) would have be due
to the above binary hadronic reactions. In addition, on the theoretical side,
thermal models have become very popular to explain the numerous
ratios of produced hadronic particles within simple thermodynamical
terms, i.e. by only specifying the temperature $T$ and the appropriate
chemical potentials $\mu_B$ and $\mu_s $ for the baryonic particles
and the strange hadronic particles
\cite{CS93,LTHSR95,BMS96,BGS98,CR00,Sp98}. The success
of this rather simple analyses strongly signals the
existence of an almost completely thermally as well as chemically
equilibrated hadronic system where the particles chemically freeze
out. (The thermodynamical properties found by the various groups
can phenomenologically easily be explained in fact by a rapidly hadronizing and
disintegrating QGP phase \cite{Sp98,St99}.)
Chemical equilibration is found to be true also for the strange
particles and especially also
for the antihyperons and multi-strange baryons.
Sometimes, though, a so called strangeness suppression factor `$\gamma _s$'
for each unit of strangeness contained in a specific hadronic particle
is introduced to slightly better account
for a common overall fitting to the ratios \cite{LTHSR95,BGS98,CR00};
this factor typically varies around $\gamma_s \approx 0.7-0.9$
and is thus close to unity.

Although the arguments at hand seem rather plausible,
it has been shown by means of sophisticated hadronic transport algorithm
like RQMD \cite{Ma89} or HSD \cite{Ge98} that at least the major amount
of produced strange particles, kaons, antikaons and Lambdas
can be understood in terms of still energetic
and non-equilibrium secondary and ternary interactions
among nucleons and already
produced mesons. This is especially true at SpS-energies \cite{Ge98},
whereas at lower AGS-energies some smaller deficiency still persists.
Only for a system close to thermal equilibrium, as was assumed
in the early calculations \cite{KMR86}, the strangeness production rates
are substantially suppressed due to the high thresholds when
considering such oversimplified initial conditions \cite{BCGEMS00}.
The conclusion that a QGP is needed to explain
the overall strangeness production seems to be considerably
weakened and thus premature if it is not to explain for the
enhanced production of the rare antihyperons and multistrange baryons!
A few phenomenological attempts to explain the more abundant production within
a hadronic transport description do exist like the color rope formation
by Sorge et al \cite{S95} or the high-dense cluster formation of
Werner and Aichelin within the VENUS code \cite{WA93}. The underlying
mechanisms, however,
have to be be considered as exotic
(like also the ad hoc dynamical formation of the QGP).
In addition, the agreement with data is quantitatively not completely
satisfying \cite{WA97b}.

In the following we present and elaborate on a convincing
argument that not binary hadronic reactions, as considered above,
but in fact multi-pionic and kaonic interactions in a thermalized hadronic gas
lead to a very fast chemical equilibration of the antihyperon
degrees of freedom. Our argument will be based on two rather
moderate assumptions:
(I) The thermally averaged annihilation cross section
for antihyperons colliding with a nucleon, i.e. $\bar{Y} + N $,
is roughly as large as the measured one for $\bar{p} + p$
or $\bar{p} + n$.
(II) At the onset for the equilibration of the antihyperons
we assume a hadronic fireball
(with thermodynamic parameters as obtained e.g.~in
\cite{CS93,LTHSR95,BMS96,BGS98,CR00,Sp98}), where the pions together with the
nucleons {\em and} the kaons are assumed to be
nearly in chemical equilibrium. As discussed above, the
abundant and early production of
kaons and antikaons can reasonably be accounted for by sophisticated
hadronic transport models without the need for a QGP.

Indeed the idea was triggered by a recent work of
Rapp and Shuryak who described the maintenance
of nearly perfect chemical equilibrium of antiprotons
together with pions and nucleons during the late stage of the expanding
hadronic fireball until thermal freezeout at rather low temperatures
of $T\approx 120 $ MeV \cite{RS00}.
They argued that the
balance between the inverse multi-pionic
channel of the reaction
\begin{equation}
\label{antip}
\bar{p} + N \, \leftrightarrow \, {n} \pi
\, \, \, ,
\end{equation}
where ${n} \approx 5-7 $ denotes the typical number of pions,
together with the strong
annihilation rate $(\Gamma_{\bar{p}})^{-1}=\tau_{\bar{p}}\approx 3$ fm/c
leads to an effective chemical potential for the antiprotons
\begin{equation}
\mu_{\bar{p} } \approx - \mu_p + {n} \mu_\pi = - \mu_B + {n} \mu_\pi
\, \, \,  ,
\end{equation}
which can account for the antiproton yield also with parameters
at thermal freeze-out. The pions are assumed here to acquire a nonvanishing
chemical potential $\mu_\pi \neq 0$ as the inelastic annihilation of pions cannot be
sustained at lower temperatures \cite{Ka90}. In any case
this is a remarkable observation
as it clearly demonstrates for special observables
- like the considered abundance
of antiprotons - the importance of kinetic multi-particle channels.
It is rather well-known that the overall $\bar{p}$ yield can be hardly
described within standard transport approaches due to the large
annihilation cross section without invoking new ad hoc assumptions or scenarios
\cite{BBBSS00}. This circumstance is then exactly due to the violation of detailed
balance when not considering the multi-particle `back-reaction' as the
most dominant source of production. Naively one might argue that the probability
of $5-7$ pions to come close in space is very low and therefore irrelevant. This however
is misleading. In fact concerning e.g.~possible changes in the total pion number the
reaction (\ref{antip}) might be neglected. For $\bar p$s (and for anti-strange baryons),
however, the situation is different since their abundance is very low. Therefore also
less likely reactions may become important. 
\begin{figure}
\centerline{\psfig{figure=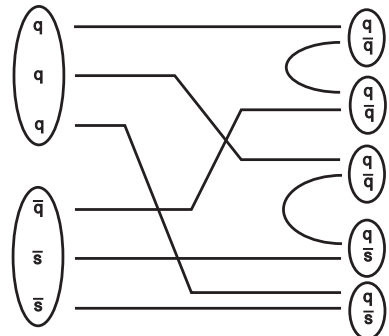,width=5cm}}
\caption{Schematic picture for 
$\bar \Xi + N \to 3 \pi  + 2  K$. 
See main text for details.} \label{fig:hyperon} 
\end{figure}

Let us consider the analogous annihilation reactions
involving one antihyperon
\begin{eqnarray}
\bar{\Lambda } + N  & \leftrightarrow & {n}_{\bar{ \Lambda }}\,  \pi  + K  \nonumber
\\[2mm]
\bar{\Xi } + N  & \leftrightarrow & {n}_{\bar{ \Xi }}\,   \pi  + 2  K
\label{antihyp}
\\[2mm]
\bar{\Omega } + N  & \leftrightarrow & {n}_{\bar{ \Omega }}\,   \pi  + 3  K \nonumber
\end{eqnarray}
or, in shorthand notation,
\begin{equation}
\label{antihyp1}
\bar{Y} + N \, \leftrightarrow \, {n}  \pi + n_Y  K
\, \, \, ,
\end{equation}
where $n_Y$ counts
the number of anti-strange quarks within the antihyperon $\bar{Y}$,
and, in direct analogy to reaction (\ref{antip}),
${n}+ n_Y$ is expected to be around $\approx 5-7 $.
A typical annihilation reaction is schematically depicted in Fig.~\ref{fig:hyperon}.
The reactions (\ref{antihyp}) are exothermic as the reaction (\ref{antip}).
As an educated guess it is plausible to assume that the
annihilation cross sections are approximately the same as for $N\bar{p}$
for the same relative momenta. In the relevant regime
of a thermal hadronic gas with temperatures of $T\approx 150 - 200 $ MeV
one has $\sigma _{p \bar{Y}\rightarrow n  \pi + n_Y K}
\approx \sigma _{p\bar{p} \rightarrow n \pi}  \approx 50 $ mb
in the relevant energy regime \cite{RS00,PDB}.
(Of course, though, the annihilation cross section
of equally charged particles like e.g. $\bar{\Xi }^+ + p$
will be Coulomb suppressed at very low relative momenta, but this is not
of much importance for the thermally averaged cross section.)
At the onset of thermalization
and chemical equilibration for all other degrees of freedom
in the hadronic fireball
the baryon density is still rather large and might exceed two
times normal nuclear matter density. For the following estimate we use
$\rho_B \approx 2 \rho_0 $ \cite{BMS96}.
One then finds for the inverse of the thermal reaction rate which as we shall see
equals the chemical
equilibration time of the antihyperonic particles
\begin{eqnarray}
(\Gamma _{\bar{Y}})^{(-1)} \, &=& \, \tau_{\bar{Y}} \, := \,
\frac{1}{\expl \sigma _{N \bar{Y}\rightarrow n  \pi + n_Y K}
v _{\bar{Y}N} \expr \rho_B } \, \nonumber \\[2mm]
&\approx & 1 - 2 \, \mbox{fm/c}
\, \, \, ,
\label{taueq}
\end{eqnarray}
which is indeed very small and much below the typical fireball lifetime
of $5-10$ fm/c. Antihyperons are forced rather immediately to local
chemical equilibrium together with the pions, kaons and nucleons!
There is no need for any exotic explanation, either hadronic
or by coalescence out of a potential QGP, to account for the thermally
and chemically equilibrated total particle number of antihyperons.

We have to note that the consideration of the above reactions (\ref{antihyp})
is not new. In fact, this has been taken into account already
in the master equations for the
strange hadronic particle densities developed by Koch et al \cite{KMR86}.
The obvious question is then why they had not obtained our present conclusion,
but much to the contrary put forward the by now famous
agenda for the antihyperons as a clear
signature of a QGP! Looking at Fig.~B3 in \cite{KMR86} they
have only considered the annihilation cross section $
\sigma_{p\bar{p}\rightarrow 5 \pi} \approx 10$ mb,
which is a factor of 5 or so smaller than the total annihilation cross section
$\sigma_{p\bar{p}\rightarrow n \pi} \approx 50$ mb \cite{PDB,RS00} in the
relevant kinematic regime. As the obtained results in \cite{KMR86}
do still show a much slower equilibration rate, a definite conclusion
cannot be given \cite{remark}.

To present our argument more definite, let us write down the dominant
contribution for the master equation of the antihyperon density
in a hadronic gas following
closely the notation in \cite{KMR86}:
\begin{eqnarray}
\label{mastera}
\frac{d}{dt} \rho _{\bar{Y}} \, &=& \,   - 
\expl \sigma _{\bar{Y}N} v _{\bar{Y}N} \expr
\left\{
\rho _{\bar{Y}} \rho_N \,  \vphantom{\sum_{n}} \right. \\
&& \left.
-  \, \sum_{{n}}
{\cal R}_{(n,n_Y)}(T,\mu_B,\mu_s) (\rho _\pi)^{{n}} (\rho _K )^{n_Y}
\right\} \, \, \, ,  \nonumber 
\end{eqnarray}
where $\rho_i$ is the density of species $i$ and
\begin{equation}
{\cal R}_{(n,n_Y)}(T,\mu_B,\mu_s) \, = \,
\frac{ \rho _{\bar{Y}}^{eq.} \rho ^{eq.}_N }
{(\rho ^{eq.}_\pi)^{{n}} (\rho ^{eq.}_K )^{n_Y}}
\end{equation}
denotes the appropriate factor for {\em assuring detailed balance} 
which depends only on the temperature and the chemical potentials.
Here we sum over all possible final number $n$ of
pions.
$\expl \sigma _{\bar{Y}N} v _{\bar{Y}N} \expr$ denotes the thermally averaged
cross section as defined in \cite{KMR86}.
As the nucleons are the most dominant
baryonic particles we take $\rho_N \approx \rho_B$ and employing~(\ref{taueq})
the master equation (\ref{mastera}) can be
brought in the more intuitive form
\begin{equation}
\label{masterc}
\frac{d}{dt} \rho _{\bar{Y}} =    
- \Gamma  _{\bar{Y}}
\rho _{\bar{Y}} \, +  \,
{\cal G}_{\bar{Y}} \, \, \, ,
\end{equation}
where $\Gamma  _{\bar{Y}} \approx 0.5-1 \, $c/fm.
If the pions, nucleons and kaons stay in thermal and chemical
equilibrium (assumption II), i.e.
\begin{equation}
  \label{eq:equil}
\rho_\pi = \rho ^{eq.}_\pi \quad , \quad  \rho_N = \rho ^{eq.}_N \quad , \quad  
\rho_K = \rho ^{eq.}_K
\end{equation}
(\ref{masterc}) simply becomes
\begin{equation}
\label{masterd}
\frac{d}{dt} \rho _{\bar{Y}} \,  = \,    - \,
\Gamma  _{\bar{Y}}
\left\{
\rho _{\bar{Y}} \, -  \,
\rho ^{eq }_{\bar{Y}}
\right\} \, \, \, .
\end{equation}
Thus the equilibration time is given by $\Gamma^{-1}_{\bar Y}$ as advocated above.

From (\ref{masterc}), (\ref{masterd}) it follows that the
multi-mesonic back-reactions, leading to a production term
${\cal G}_{\bar{Y}}$, are necessary to achieve and to further maintain chemical
equilibrium of antihyperons with pions, kaons and nucleons.
This multi-mesonic source of production of antihyperons is a consequence
of detailed balance and, as the rate $\Gamma_{\bar{Y}}$ is
indeed very large, this is the by far most dominant source
compared to the binary production channels (\ref{sproda}) and (\ref{sprodb}).
Note that as ${\cal G}_{\bar{Y}}
= \Gamma_{\bar{Y}} \cdot \rho ^{eq }_{\bar{Y}}$,
the overall production rate ${\cal G}_{\bar{Y}}$ is still a very small number.
On the other hand, these multi-particle reactions cannot be handled
within the present transport codes and are thus completely neglected
(- sometimes with the `excuse' that ${\cal G}_{\bar{p}}$ or
${\cal G}_{\bar{Y}}$
would be overwhelming largely suppressed by multi-particle phase space).
From our discussion it is obvious that
the production of antiprotons \cite{RS00} and antihyperons cannot be
addressed by these approaches. Nonetheless as we have shown there is a simple non-exotic
explanation for the $\bar Y$ abundances in a purely hadronic scenario.

If, as presented in some of the thermal and chemical models,
a strangeness suppression factor $\gamma_s $ for each unit of strangeness
is introduced \cite{LTHSR95,BGS98,CR00}, we only have to replace in (\ref{eq:equil})
and accordingly in (\ref{masterd}) 
\begin{equation}
\label{semieq}
\rho ^{eq.}_K \rightarrow
\rho_K = \gamma_s \rho ^{eq.}_K  \, \, \,
\stackrel{(\ref{masterd})}{\Longrightarrow }
\, \, \,
\rho _{\bar{Y}}^{eq} \rightarrow
\rho _{\bar{Y}} =
(\gamma_s )^{n_Y} \rho _{\bar{Y}}^{eq.}
\, \, \, .
\end{equation}
Accordingly, the antihyperon density would aquire an additional factor
$(\gamma_s )^{n_Y}$ compared to chemical equilibrium value from the
stationary limit of the master equation, if
the kaon number is not fully saturated. This is consistent
with the employed phenomenological prescription
in \cite{LTHSR95,BGS98,CR00}.

As a final comment we briefly consider
the situation of antihyperon production in
relativistic nucleus-nucleus-collisions at the AGS
and at RHIC:
According to the thermal models the deduced temperatures
at the AGS-energies are lower
and the obtained baryon densities are higher \cite{BMS96,CR00}.
The latter would imply that according to (\ref{taueq})
the antihyperon chemical equilibration time $ \tau_{\bar{Y}} $
becomes even smaller.
As the pions and to some good extent also the kaons are found
to be in chemical equilibrium \cite{BMS96,CR00}, our argument should perfectly
apply!
Indeed, the E859 Collaboration has measured the $\bar{\Lambda }/\bar{p} $
ratio in Si+Au at 14.6 AGeV and had reported a large value
$\bar{\Lambda }/\bar{p} = 2.9 \pm 0.9 \pm 0.5 $ for some central
rapidity window \cite{E859}. This would favor a scenario of nearly
chemically saturated strange and nonstrange antibaryon
populations at particle freeze-out. More precise experimental data
as well as a theoretical analysis along the line of thermal models
would be most welcome.
The situation is different at RHIC. The typical baryon densities
at midrapidity should be rather small, if not negligible.
Hence, our considered equilibration time for the antihyperons
in a baryon dilute hadron gas should be much larger.
In this case, if the fireball does not stay too long in the
late stage hadronic phase, chemical equilibrium of the antihyperons
cannot be reached or maintained, if there is no other source
of their production.

\acknowledgments

The authors thank U.~Mosel for stimulating conversation
when discussing the ideas and line of arguments.





\begin{references}
\bibitem{remark}
Unfortunately, detailed numbers
for the thermally averaged cross sections were not explicitly specified in \cite{KMR86}.
\end{references}
\end{document}